\documentclass[aps,preprint,amsmath,amssymb]{revtex4}
\usepackage{amsmath,amssymb}
\usepackage{graphicx}
\usepackage{booktabs}
\usepackage[usenames, dvipsnames]{color}
\usepackage{multirow}
\usepackage{setspace}
\usepackage{rotating}
\usepackage[ pdftex, plainpages = false, pdfpagelabels, 
                 pdfpagelayout = useoutlines,
                 bookmarks,
                 bookmarksopen = true,
                 bookmarksnumbered = true,
                 breaklinks = true,
                 linktocpage=all,
                 pagebackref=false,
                 colorlinks = true,
                 linkcolor = black,
                 urlcolor  = blue,
                 citecolor = black,
                 anchorcolor = green,
                 hyperindex = true,
                 hyperfigures
                 ]{hyperref} 

\begin{document}
\setcounter{page}{0}

\title{\href{http://www.necsi.edu/research/military/cyber/}{Principles
    of Security: Human, Cyber, and Biological}} 
\author{Blake C.\ Stacey and
        \href{http://necsi.edu/faculty/bar-yam.html}{Yaneer Bar-\!Yam}}
\affiliation{\href{http://www.necsi.edu}{New England
    Complex Systems Institute} \\ 238 Main St.\ Suite 319 Cambridge MA
  02142, USA \vspace{2ex}}

\date{June 1, 2008 / public February 28, 2013}

\begin{abstract}
Cybersecurity attacks are a major and increasing burden to economic
and social systems globally. Here we analyze the principles of
security in different domains and demonstrate an architectural flaw in
current cybersecurity. Cybersecurity is inherently weak because it is
missing the ability to defend the overall system instead of individual
computers. The current architecture enables all nodes in the computer
network to communicate transparently with one another, so security
would require protecting every computer in the network from all
possible attacks. In contrast, other systems depend on system-wide
protections. In providing conventional security, police patrol
neighborhoods and the military secures borders, rather than defending
each individual household. Likewise, in biology, the immune system
provides security against viruses and bacteria using primarily action
at the skin, membranes, and blood, rather than requiring each cell to
defend itself. We propose applying these same principles to address
the cybersecurity challenge. This will require: (a) Enabling pervasive
distribution of self-propagating securityware and creating a developer
community for such securityware, and (b) Modifying the protocols of
internet routers to accommodate adaptive security software that would
regulate internet traffic. The analysis of the immune system
architecture provides many other principles that should be applied to
cybersecurity. Among these principles is a careful interplay of
detection and action that includes evolutionary improvement. However,
achieving significant security gains by applying these principles
depends strongly on remedying the underlying architectural
limitations.
\end{abstract}

\maketitle

\begin{equation*}
\begin{array}{lccccccr}
\includegraphics[height=5cm]{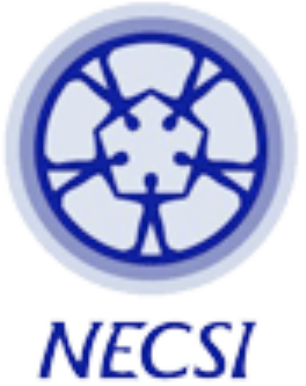} & \qquad & \qquad & \qquad & \qquad & \qquad & \qquad &   \qquad \includegraphics[height=5.5cm]{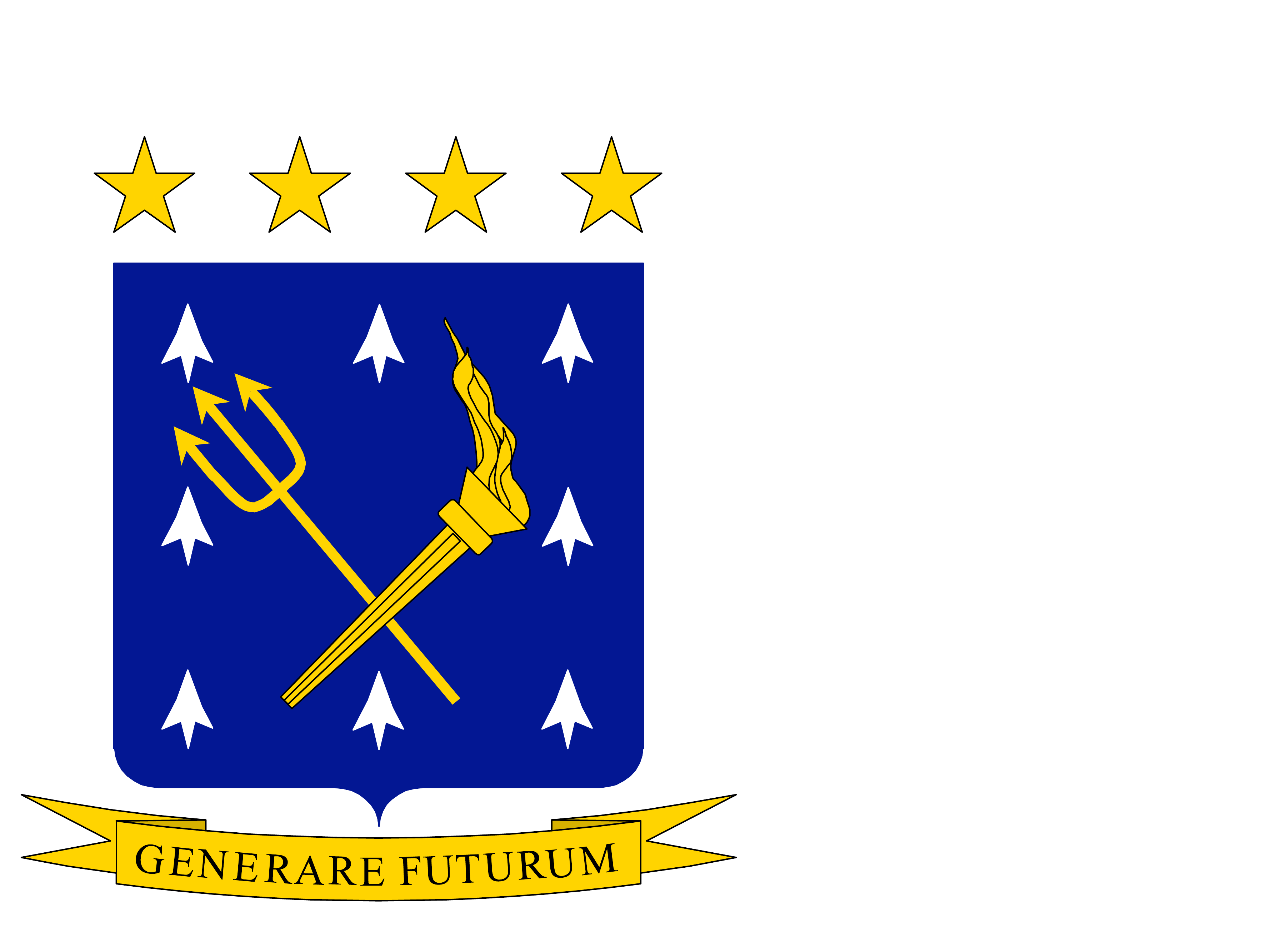} \\
\end{array}
\end{equation*}

\vspace{1.5cm}

\begin{flushright}
\Large{June 1, 2008}
\end{flushright}
\large

\Large 
\textbf{ Principles of Security: Human, Cyber and Biological  }
\bigskip

\vspace{1cm}

\large
Blake Stacey and Yaneer Bar-Yam
\normalsize
\vspace{.2cm}
 {\obeylines 
New England Complex Systems Institute
24 Mt. Auburn St. 
Cambridge, MA 02139
}

\vspace{3cm}

{\obeylines 
\normalsize
Reported to
\large
William G. Glenney, IV
\vspace{.2cm}
\normalsize
Chief of Naval Operations Strategic Studies Group}

\normalsize 
\thispagestyle{empty}
\vfill\eject

\setcounter{page}{1}
\pagenumbering{roman}

\Large
\noindent Preface
\normalsize

\vspace{.5 cm}

In developing revolutionary concepts of Naval Warfare,\cite{SSG97,SSG98,SSG99,SSG00,SSG01,SSG02,SSG03,SSG04,SSG05,SSG06} the Chief of Naval Operations Strategic Studies Group (SSG) has anticipated concepts that are being now adopted widely. Among these concepts are Network Centric Warfare and FORCEnet. \cite{SSG97,SSG01,NCW}  Modern technology enables a new focus on the interactions and relationships of action comprising a network of human beings and technology as central to advantage in military conflict. More significantly, the challenges of 21st century warfare require significantly greater levels of capability for which networks appear to be well suited, and even necessary.

In support of the SSG, Yaneer Bar-Yam, president of the New England Complex Systems Institute (http://www.necsi.edu), has provided a fundamental scientific basis using multiscale representations for analysis of the capabilities of organizations. [12-21] His analysis characterizes the capabilities and limitations of traditional and networked military organizations.

Through periodic lectures at the SSG beginning in January 2000 that have informed SSG reports, and through papers \cite{YBSSG1,YBSSG2,YBSSG3,YBSSG4}  addressing specific questions in military force organization and transformation, a basis for understanding is being developed of modern military conflict that enables generalization from experience gained to new and novel military responsibilities. 

One of the central insights provided by multiscale representations is that there are distinct types of networks that are relevant to military operations. \cite{MTW,YBSSG3,YBSSG4} One of these is a network of agents that are individually capable of action, e.g. warfighters in a battlespace. A second is a network of decision makers that gathers information widely and makes decisions collectively about highly specific actions to perform, where the action is then executed, perhaps by a large and centrally directed force. The former can be considered in analogy to the human immune system where the individual agents are the blood cells that battle harmful agents such as infections within the human body. The second can be considered in analogy to the human neuro-muscular system where the individual agents of the network are the nerve cells whose collective decision making ability far exceeds that of a single cell, and the decision making process results in the actions performed by the muscles. 

Distributed networks are particularly needed when facing enemies that are themselves distributed networks---as is apparent in counter terrorism and in cyber security. Both are growing challenges. The attached report focuses on cyber security. Today, cyber exploits---spam, malware, denial of service, and security breaches---are widespread. As cyber space becomes an increasingly integral part of the functioning of all human activities security in this domain becomes an increasingly critical and unmet challenge. This report presents a framework for understanding the inherent limitations  of existing cyber security and how to fundamentally improve its capabilities.

\vfill\eject

\Large
\noindent Executive Summary
\normalsize

\vspace{.5 cm}

Increasing global interdependence has resulted in distributed human terror networks and cyber security challenges. Actions by and on networks can result in major damage even though individual agents causing it are difficult to locate or identify. Traditional methods of centralized control and local response cannot meet these challenges due to the fundamental constraints on distributed information gathering, detection and coordinated action. Distributed networked security systems are necessary for effective response. Rather than addressing specific challenges of today, the vulnerability caused by global communication and transportation must be met by creating security systems with necessary capabilities to meet a wide range of challenges that are possible in these systems. With traditional approaches to security cyber systems are particularly at risk given the rapidity of action and necessary response.

Security challenges follow certain general patterns. Recognizing the principles of effective security response provides guidance to meeting new high complexity challenges in global terror networks and cyber security. Key principles can be found from analyzing security systems in other domains. In this paper we analyze the biological immune system which is responsible for analogous security problems within human physiology. The immune system has evolved a dispersed distributed-control system that can successfully respond to generic and novel threats and  can distinguish threat agents from self and attack threats/invaders with necessary force. 

The human immune system consists of billions of cells that coordinate response to security challenges in the human body. The activity of the immune system cannot be understood as a centrally managed process, but rather as arising from a large number of local communications among specialized components to achieve an emergent response. It achieves high success rates, is robust, scalable, flexible, and is generally capable of distinguishing self from non-self in actions. It dynamically refines its ability to detect pathogens by evolutionary selection. 

Actions of the immune system can be divided into three layers. The first layer consists of barriers between security domains, including the skin and membranes between compartments of the body. The second layer consists of responses to damage of the system affecting many cells, including repairing barriers and tissues. The third layer, the adaptive immune system, responds to the finest scale challenges, detection and response to cells or molecules distributed through the system. Of particular concern are the viruses and bacteria that can replicate rapidly from a small to large number. All layers of the immune system have analogs in human and cyber security. The most difficult to understand and therefore the focus of our attention is the adaptive immune system.  

The central capabilities of the adaptive immune system are detection and action, whose generalized application must be found in any security system. Detection is achieved by mapping the space of observed structures or behaviors onto a reduced set of possibilities, which are partitioned into threat and non-threat, i.e. non-self and self. When a match is found in the environment to a non-self template, an action is triggered. Detection and action are separated functions of components of the system, requiring careful regulatory interplay between them to enable action without damage to self, corresponding to collateral damage. Specific details of communication protocols and interplay of detection and action agents reveal the strategies the immune system employs in a system wide response whose detection ability increases rapidly in order to eliminate even individual molecular or cellular harmful agents. This requires the ability to replicate and distribute the detection mechanisms and action response rapidly throughout the system. The rapid development of improved response involves progressive refinement of detection by a process corresponding to evolutionary selection, similar to breeding of desired features in agriculture. Improved detection templates developed for a particular threat are distributed throughout the system to enable effective local response.

The relevance of the immune system model to cyber security arose when computer communication systems transitioned from infrequent and low bandwidth processes to increasingly pervasive, persistent and high speed networks. The importance of security is growing as the Internet is used not just as an adjunct to activities for communication about them, but rather as a necessary component of a wide range of economic and social functions. The need for security is apparent in the large impact of spam, spyware, phishing, zombie networks, denial of service attacks, internet fraud, identify theft, and breaches of high security systems. The underlying cause of the need for security is the transition of the Internet connected computers to behavior as a tightly linked system analogous to a multicellular organism.

Existing cyber security systems have parallels to the immune system but these parallels are incomplete. Corresponding to the barriers found in the first level of security are firewalls and separation of distinct networks, e.g. ATM and bank transaction networks where these are separated from the Internet. The second layer of security consists of response to widespread exploits, including Domain Name Server Black Lists (DNSBLs) for blocking spam from their sources. The third layer of security includes virus scanners and e-mail filters that detect malware or spam on an individual computer.  Detection makes use of a characterization of programs or e-mail by features (bit strings and logical operations on these bit strings) that provide signatures similar to the detection templates of the immune system. Considering the correspondence with the immune system reveals that the response to detection in cyber security is not pervasive throughout the system.

Our analysis of the immune response identifies the principles of successful response. While there are similarities to the immune system response in cyber security, we find two major gaps in the architecture of the Internet. Without addressing these issues, improved cyber security will be difficult or impossible to attain. First, there is no mechanism for distribution of a response throughout the system. While individuals can subscribe to security systems that are distributed to some degree, it is optional. Moreover, local information about detection is not redistributed throughout the system. Note that malware does have the capability of replication and distribution. The absence of parallel capabilities in security agents enables attackers an inherent advantage over security efforts. Second, the existing security system is not a collective security system in that it does not protect the Internet but rather protects individual components of the Internet. The ability of any part of the Internet to send messages to any other part of the Internet without encountering security systems implies that weakest elements can be attacked, compromised, or controlled to enable progressively larger infestation of the system. Conversely, the inability to mount a pervasive defense is a fundamental limitation on security which is diametrically counter to the corresponding processes of immune system response. Without a collective security system, the only methods for progress are to harden each element of the Internet, a much larger security burden destined to be ineffective. Specifically, the only means are to improve the security of the operating system on each computer. Even so, computers that are operated by individuals that desire to cause harm would continue to be security risks. The implications of these gaps in the security system are that there is no mechanism within cyberspace for security actions to counter-attack the sources of attacks. Moreover, there is no mechanism for blocking their attacks at point of entry into the Internet rather than at point of attack at another node. 

Addressing cyber security would require either or both: (a) Making pervasive distribution of self-propagating but non-destructive security ware acceptable and create a developer community for such security ware. (b) Modifying the protocols of internet routers to accommodate adaptive security software that would regulate internet traffic of other kinds and self-regulate. These modifications would alter the perspective of the "rights" of the Internet, the right of transmission and the right of any node to communicate to any other node of the system. An effective security system requires that this right be limited, as best as possible, to those who do not cause damage to the system.

The analysis of the immune system architecture provides many other principles of security that can be applied to cyber security. Among these principles is a careful interplay of detection and action that includes evolutionary improvement of the detection and response capability. However, any advances in applying such principles will have minimal impact as long as the underlying architectural limitations persist.

\vfill\eject

\tableofcontents
\vfill\eject

\setcounter{page}{1}
\pagenumbering{arabic}
\section{Overview on Security}
The current strategies used in human and cyber security are not
capable of handling threats in our increasingly interdependent
world. Challenges in human security are changing through global terror
networks.  Cyber security, by virtue of its rapid and hidden processes
is arguably an even greater challenge that is poorly met by existing
systems. Severe exploits and massive system burden from actors in
cyberspace are common today. Were this extent of malware (viruses,
trojans, spyware) and spam found at the human level, it would be
considered a major breakdown of a social system.

The demands of addressing current challenges in human and cyber
security are motivating the development of fundamentally new
approaches. An essential feature of new challenges is their
distributed nature. Global transportation and communication systems
enable distributed groups of individuals to cause major physical or
informational damage, elevating the global challenge of maintaining
security at any location.  On the one hand, traditional police forces
with solely local authority cannot respond to global relationships and
associations. On the other hand, the many possible actions of diverse
forms can overwhelm centralized responses due to the inability to
gather and process information, determine courses of local action, and
distribute control messages appropriately. The distributed nature of
these challenges demands a distributed, and correspondingly complex,
response.\cite{MTW}

Traditional security revolves around localized agents of various
sizes, from individual criminals to national armies. Such agents
manifest in the local damage they cause. To combat such agents, the
source of the damage must be identified and an appropriate local
response mounted. In general, the possible damage is limited to the
scope of the agent. Small actions are harder to detect, but are
commensurately responsible for less damage. Larger actions responsible
for larger-scale damage are easier to observe by virtue of their
size. From this perspective, coordinated, distributed actions present
a new challenge in that any one of them is difficult to detect, the
impact of any one agent may be in a different location, and the
collective effect of multiple agents can be large.  Moreover, an
additional universal aspect of these challenges is their ability to
proliferate so that a small scale challenge can grow by recruitment to
a large scale over a short period of time.

An essential aspect of distributed, coordinated action is the
availability of necessary communication and transportation
mechanisms. It is essential to recognize that modern social and
technological changes create the opportunities for communication,
transportation and recruitment by virtue of the structures that are
developed for the functioning of society itself.  Changes in global
connectivity are driving vulnerability to security challenges that are
inherent in the structure of the system.

Recognizing that security challenges stem from changes in social
connectivity is important. Otherwise we may be fooled into addressing
only specific current challenges, and fail to meet the next challenge
that arises.  Solving the current challenge will not eliminate the
general vulnerability. Instead, it is necessary to create systems that
are able to address challenges of various kinds that surely will arise
in the future.

The changes in global communication and transportation systems are
easily recognized in the human sphere given the increasing ease and
frequency of global transportation and communication among
people. However, it is even more dramatic in the cyber space context
where changes can be more rapid and the time scale of response may be
shorter.

A useful paradigm for the study of distributed response security
systems is the biological immune system. The immune system is the
biological system responsible for internal security. The analogy
between biological attackers and computer attackers is well
established in the notion of ``computer viruses.''\cite{ref:cohen}
Still, the key to developing effective systems of security is
understanding the principles of immune system function so they can be
generalized. In this paper, we frame the essential issues to guide a
more general approach to using the structure and processes of the
immune system to inform distributed human and information technology
security. We focus on the part of the immune system designed to
address the most complex distributed challenges corresponding to the
highly complex distributed challenges in human and cyber security.  In
this context we detail how the distributed structure and the key
interactions between agents enables the remarkable level of success
achieved in a demanding security environment. This success is measured
by the ability to eliminate to the last one adversaries that are
capable of rapid proliferation if left unchecked.

This paper is organized upon the concept that identifying principles
from one security system can serve as a foundation for
modeling/organizing other security systems with comparable complexity.
Understanding principles of security from the immune system is not the
same as making an analogy between biological and social or
technological systems. Principles embody correspondences that reflect
essential logical or mathematical relationships between elements,
structures and behaviors.  While we do not provide here the formal
correspondence, we strive to make the nature of the formal
relationships clear in the presentation.

Our primary conclusions are that the existing structure of the
Internet does not allow for a security system that is able to address
its security challenges. This architectural problem supersedes all
specifics, as it is necessary to address the underlying architectural
problems in order to enable implementation of effective security.

In order to establish the fundamental gap in the ability of the
current security systems we explain the functioning of the biological
immune systems solution to its own security problem. We then discuss
the mechanisms that have been used to address cyber security, why they
fail, and what changes are necessary to enable success.

\section{The Immune System}

\subsection{Introduction to the immune system}

Human beings possess an immune system responsible for the well-being
of trillions of cells, itself comprised of billions of
interacting pieces, with no single central focus of control. The
immune system is a medically significant arena of study, as it
provides our defense against diseases, and failures of its operation
lead to auto-immune disorders. It also provides an important context
for developing our understanding of universal principles that apply
across diverse complex systems including social and technological ones.

The immune system can be understood as a system with emergent
response.  The many components of the immune system function by
collectively responding to challenges.  The natural scale of its
response locally may involve only a few cells, throughout the system
it involves many cells responding to potentially distributed
challenges in a coordinated fashion. When necessary, the response in
any one location can increase dramatically to achieve a macroscopic
response as is found in visible infections.  Collective actions are
distributed among multiple cells, achieving a balance where each cell
is important but generally not essential.  This emergent behavior
cannot be understood by description of individual agents. It must be
understood by the coupling of individual cell actions through
inter-cellular interaction.

The immune system has properties that are desirable in any security
system: first, it is {\em robust and resilient,} as individual
components of the system can be removed without compromising the
functionality of the whole.  Second, the system is naturally {\em
  scalable} without substantial modification.  Third, it is {\em
  flexible,} often able to cope with pathogens which have never been
seen before.  Fourth, it displays {\em specialization,} with different
cell types performing the different functions necessary to provide
system functionality.  Fifth, it is able to distinguish {\em self from
  nonself,} even when the molecules it produces during its normal
affairs are themselves novel chemical structures.

The immune system exploits the dynamics of evolution to achieve its
remarkable success. Evolution is a central process in the formation of
any complex system and is fundamentally necessary to achieve success
in complex tasks. \cite{MTW} Evolution enables improvement of a system
to face challenges that cannot be anticipated.  Understanding how this
works is therefore necessary for our ability to address many
challenges.  In particular, it is important to understand that not
only is the immune system itself a product of evolution, but it also
applies evolution to address specific immune responses.  This latter
evolution takes place within a single organism.

\subsection{Immune system architecture}

The biological immune system guards against infection in multiple
ways, providing a ``layered defense'' with increasing specificity.
Pathogens which penetrate system barriers ({\em e.g.,} skin) are
challenged by the {\em innate immune system,} and the {\em adaptive
  immune system.}  These mechanisms are constantly active, meeting a
dynamically changing and large set of challenges at all
times. Significant illness occurs only when they fail.

The first layer of security consists of the barriers themselves.
Barriers distinguish the self from other at the large scale. They
separate space to that controlled and not controlled, or subject to
different levels of control. The immune system includes the skin and
other membranes that separate specific compartments of the body. The
skin is the boundary that separates the internal space where the
immune system controls the fine scale security, from the space outside
the system which is not subject to security.

The second layer of the immune system is designed to meet failures of
the first layer. This includes disruptions of the barrier itself,
repairing the skin, and responding to challenges that occur when there
are significant breaches of the skin. A principle of innate immunity
is to characterize damage and threats that are of intermediate scale
--- involving many cells --- and therefore generic features of large
classes of hazards rather than specific to the attacker types.  These
include response to blood loss (clotting) or to tissue damage
(inflamation).

The third layer of the immune system addresses the finest scale
challenges. In particular, the role of adaptive immunity is to
identify specific individual threats that do not trigger the innate
immune response: individual molecules, cells or loosely associated
groups of molecules or cells distributed through the system.

While there are many potential causes of harm that the immune system
guards against, there are two that are of particular note because they
are able to self-replicate and thus require a response that both can
address a large number of such attackers and yet detect a very small
number, even down to a single individual. There are two primary types
of such foreign agents: molecular attackers, viruses, and cellular
attackers, commonly bacteria but also other replicating single celled
organisms.

\subsection{Specific fine scale (adaptive) response}

Our focus is on the adaptive immune system whose role is to identify
and attack specific intruders or self cells that change behavior to become
a threat to survival. Any security system must perform certain key tasks.

\begin{enumerate}

\item {\bf Detection.}  Hostile elements introduced into the system
or arising within the system must be noted, distinguished from 
non-hostile elements and identified before action can be taken.

\item {\bf Action.}  Once a threat has been identified
and characterized, a response must be mounted which is appropriate to
both the quantity and the nature of the threat.

\end{enumerate}

We now address these points in more detail.

\section{Adaptive Immune Response}

\subsection{Identification}
The essence of recognizing a threat is distinguishing signatures of
its structure or behavior. The identification of a foreign ``antigen''
in the immune system is analogous to distinguishing undesirable from
desirable elements in any system. While this is a commonly necessary
task, this form of pattern recognition is fundamentally
difficult. Understanding how it is accomplished is key to the design
of any security system.

Central to identification is a way to map any structure or behavior
that is to be characterized onto a smaller and better defined set of
``possibilities.'' This set of possibilities is then partitioned,
perhaps dynamically, into a set that are considered a threat and those
that are not a threat.

The immune system performs this distinction by starting from a large
set of prototypes within the space of possibilities, and rejects from
this set those that are matches to components of self.  The remaining
set are used for identification of antigens. Also, perhaps, damage
caused may be used as evidence. When an antigen is detected, action is
initiated, and the prototypes are refined by evolutionary processes to
achieve better matching, and thus more rapid response.

\subsection{Capacity for action}

The identification of the presence of a threat must trigger a response
that will be effective in meeting that threat. The response could be
of various kinds, but it must modify to render harmless, or eliminate
the threat from the system.

In a distributed system, such as the immune system, an appropriate
level of response must be recruited from nearby or from far away
depending on the size of the response needed. Once the existence of a
threat has been identified, the ongoing need for detection should be
simplified for the responding agents. This enables action to be
performed by components that are not as capable of performing
differentiation by themselves, but are instead specialized for action
against a threat.

The existence of responders who are then less capable of
discrimination, however, increases the likelihood of actions that are
against self rather than against threats. This challenge must be met
by careful regulatory processes for triggering action. The need to
identify threats, and then selectively trigger action requires
balanced protocols to safeguard self while defeating threats.
Examples of this issue in human security include the problems of
``friendly fire'' and ``collateral damage'' in warfare, which
constitute harm to responders and to bystanders respectively.  While
these terms imply unintended harm, we can also include in this
category intentional actions by individual agents that are not
consistent with collective goals, such as war crimes.  There are two
types of methods for avoiding self-damage.  The first are within an
individual agent, analogous to intention of the individual, and the
second involve regulatory interactions between agents at the time of
action.

In computer security, similar issues of self-inflicted damage today
include issues of ``false positives'' such as the blocking of desired
e-mails as spam, or blocking of legitimate users from accessing a
system. The extra effort involved in forgotten or mistyped passwords
and password systems can be included as well.

Multiple types of cell and multiple cells of each type are involved in
each action of the immune system. Biological and biochemical details
will be discussed in the Appendix; for the moment, we note that the
ability to spread messages and agents throughout the body via the
circulatory system allows initial signals to provoke a systemic
reaction to resolve a challenge.  This systemic reaction is not just a
call for others to respond to a particular place, but a pervasive
distribution throughout the system of the ability to respond to
attackers that can appear in a distributed fashion and replicate.

\subsection{Evolution as method}

The central principle of evolution is the ``non-random survival of
randomly varying replicators.'' \cite{Lewontin,Dawkins} In each
generation, the replicators which are better able to survive are able
to leave more offspring, and so advantageous traits can spread
throughout the population.  If the environment changes, the mixture of
traits within the population will change in response; populations
which cannot adapt in this fashion will quite generally be supplanted
by those who can.  We emphasize that the variation that takes place in
the traits through, e.g. mutations, is random, while the selection
which acts upon replicators has a much more deterministic character.

Selection creates a functionality in the replicators over time, where
that functionality can be defined by the ability to meet the criteria
of selection. By arranging criteria of selection to serve a particular
purpose, the evolutionary process can serve to induce that purpose in
the replicators.

Such a directed evolutionary process has been variously used both
historically and today.  The breeding of animals is an artificial way
of applying selection criteria to serve a purpose that human beings
determine.\cite{Darwin} In technological applications, so called
genetic or evolutionary algorithms,\cite{ref:mitchell} specify an
artificial measure of ``fitness'' of a computer-based replicator and
apply random variation and selection to improve the replicators in
meeting that measure. In this context the process of evolution can be
considered a form of optimization.  Still, computer based evolutionary
processes only incorporate some features of the evolutionary process
in nature.\cite{MTW}

If the goal of a system is relatively easy to specify but the path to
achieve that goal is unclear, then a properly implemented 
evolutionary process is an approach that can be effective.

The process of evolution is used by the immune system to
improve the most information intensive aspect of the response
process, detection.  This process requires effective use of
the existence of some detection to initiate the improvement
process so that progressive improvement of detection is
possible. Subsequently, improved detection accelerates the
ability of the system to respond. This requires rapid distribution
of the improved capabilities throughout the system through 
global dissemination. 

\section{The Cyber Security Challenge}

When early ``mainframe'' computers were replaced by personal
computers, an individual computer still remained largely isolated,
interacting with its owner. Programs or data were transferred via
floppy disk or, sometimes, over telephone lines.  As networking became
ubiquitous, the rate of ongoing interactions created a global system
in which most of the communications are not observed directly by human
beings and individual human response times are insufficient to monitor
the dynamics of the communications. Moreover, the extent and rate of
communication continues to grow.

The widespread use of the Internet for communication and commerce has
increased the need for cyber security.\cite{ref:nssc} It is important
to recognize the change from a system that is an adjunct and a
convenient substitute to more conventional communications, to when
such interactions are an integral part of society. As commerce is
increasingly done through the Internet, the disruption of economic
activity through the disruption of such communications becomes
increasingly important. The next stage of development, already
occurring in many contexts, is the integration of the Internet into
response systems. Thus the operation of almost every organization is
undergoing a transition from isolated computer stations, to
communications, to real time interactions, to integrating essential
responses through the network. Increasingly, individual actions
require information from multiple sources distributed through the
network, and actions themselves become distributed among multiple
individuals interacting through the network.  Potential disruptions of
the network system increase in impact as this occurs.

For example, computers introduced into medical operations might
first be used for tracking appointments and keeping financial records.
Then they might be used for sending prescriptions from physician
to pharmacy. Third, they can be used for real time monitoring of
procedures. Finally, they can be used for remote controlling of
procedures. Once the final stage is reached, the disruption of
service has real time impacts not only on the communication about
the procedure but the procedure itself.

The need for security has become apparent in the success of Internet
fraud, including breaches of high security systems and theft of
personal records.  The extent and variety of ``cyber'' actions has
increased with the ubiquity of spam, spyware, phishing, zombie
networks, denial of service attacks, etc.  A spam-blocking service
reports over 4.7 billion spam messages intercepted since November
2005, almost ten times the amount of legitimate traffic over that
period.\cite{ref:akismet} This includes only the spam which was
intercepted.  Many of the spam messages advertise fraudulent products
or otherwise attempt to defraud their readers; they contain links to
unlawful sites---which also serves to skew search engines like Google
in their favor; and they often originate from otherwise legitimate
computers whose security has been compromised.

The need for the biological immune system arose when single celled organisms
evolved to form multicellular organisms. As with the change in human
social and cyber connectivity, the connectivity in multicellular 
organisms leads to a collective vulnerability and a need for a system to
guard against it.

\section{Current Cyber Security}

There are a number of generally used cyber security systems that are
parallel in some way to the immune system operations. In addition,
there are specific efforts to adapt concepts from the immune system
for cyber
security.\cite{ref:forrest2007,ref:kephart,ref:forrest1994,ref:somayaji,ref:kim}

\subsection{Layered defense}

The first layer consisting of barriers in cyber security includes
firewalls and the separation of distinct networks for e.g. ATMs and
bank transactions. These security systems prevent malware from
entering a system as skin protects an organism.  Barriers within a
system, just as membranes within the body, generally are
semi-permeable, using mechanisms to differentiate what can pass. In
cyber security these include password authentication and S/Key
challenges.

The second layer of cyber security includes detection of exploits and
generic responses to them. This includes Domain Name Server Black
Lists (DNSBL) often called RBLs (Real-time Blackhole Lists). These are
services that gather and provide lists of IPs that are sources of spam
and other malware. Institutional mail servers can automatically
implement policies that use this information to block domains of the
Internet that are sources of spam. The sources of spam may include
servers set up for this purpose, or zombies which are computers that
have been compromised by malware so that they transmit spam and
malware on behalf of others. In effect, zombies are the analog of
virus infected cells that become factories for viruses and other
pathogens. The large number of these exploits today results in a
response which is akin to a generic immunity response.

The third layer of cyber security includes virus scanners and e-mail
filters are the analogs of the adaptive immune
system.\cite{ref:forrest2007} These applications search programs
stored on disk or incoming e-mail messages for signatures of malware
and spam. If the detection system is not specific enough, programs
that are valid, and e-mail that is valid are rejected. Alternatively,
malware or spam may not be rejected. The desired versus undesired
categories are analogous to the discrimination of ``self'' from
``other'' in the adaptive immune system. Where self consists of
legitimate software and desired e-mail, and other is the malware
(virus, Trojan Horse, etc.) which would compromise the system and
spam. The existence of false-positives and false-negatives that
misidentify whether the spam or malware are legitimate is similar to
errors of classification in the immune system as well.

\subsection{Detection}

A computer immune system must detect both known and hitherto unknown
viruses or spam. For this purpose a program fragment, or small piece
of data from a larger set can be used as a detection template. This
extracted data can be compared with correspondingly extracted data
from a virus or spam.  The latter is known as the ``signature'' of the
virus or spam. Finding the ``signature'' within a piece of software
indicates that the software has been infected.  Various signatures can
be constructed based upon procedures specified by individuals
(heuristic rules), or statistical pattern detection (Bayesian
filtering), and collaborative identification (when voluntary human
communities manually specify spam signatures that are
shared).\cite{SpamAssassin}

Some detection systems are local in that the software itself learns
from labeling by the user what is spam and what isn't. In this case 
the user manually identifies spam and non-spam, and signatures
are extracted automatically that differentiate between the spam and 
non-spam by the software.\cite{Bayesian} Others are
centrally directed by service providers who provide the software
and revise it to add templates for new malware.\cite{mcafee, norton, avg}
Revisions arise after reports due to the detection of the virus. Such
detection occurs when individuals observe activity of processes on 
computers outside of normal operations, or of damage due to such 
processes.

Similar issues arise at the level of computer network operation
security.\cite{ref:kim} Such security systems operate at the level of
the pattern of traffic rather than the content of individual messages.
To detect a deviant computer system that may be the source of other
attacks, a pattern detection system has a representation of the types
of patterns that can arise. Among these a set of ``self-patterns'' is
created, representing the legitimate ways in which traffic can flow
amongst the computers of a Local Area Network (LAN). Abnormal traffic,
such as a computer suddenly sending thousands of e-mail messages to
the external Internet, is a ``non-self pattern'' and is considered a
sign of infection (in this example, the computer may have been
co-opted by a spammer and used as a ``zombie'' to spread spam and
malware).\cite{ref:hofmeyr,ref:dasgupta}

\section{Why Current Cybersecurity Cannot Work}

The examples of Internet based security suggest that existing systems
have some similarities with immune system operations. Still, they do
not capture the dynamics of communication and interplay of detection
and action that should provide better security and better
self-protection. These limitations include the manner of detection and
sharing of signatures of malware (local, centralized and limited
distribution systems), as well as the limits in implementing actions
to prevent or stop attacks and exploits.

Indeed, while cyber security systems today provide some protection for
malware and spam, the ongoing presence of large volumes of spam and
malware, and exploits, suggests that the existing protections are too
limited in their abilities and greater attention to the principles of
security as embedded in immune system operations would give rise to
improved outcomes.

There are two fundamental reasons that the current approaches to
cybersecurity cannot work effectively.

First: There is no mechanism for rapid pervasive distribution of
security processes that can respond to new types of malware or spam.
The voluntary use of community or centrally generated malware or spam
signatures, while widespread, is not pervasive. One way to understand
the ineffectiveness of security distribution is to compare the
distribution of security with that of malware. Malware is much more
pervasively and rapidly distributed than the security that is designed
to guard against it. In the immune system, viruses can be considered
to be analogs of antibodies. Both are able to replicate rapidly using
cellular mechanisms and both are able to attack cells or other
molecules. This correspondence makes viruses and antibodies have
similar capabilities. Furthermore, T~cells are similar to bacteria in
their ability to replicate and attack other cells. This correspondence
of defenders and attackers implies that there are no major gaps in the
capabilities of the attackers as compared to the defenders. By
contrast, there is no defensive analog of malware, in that the
anti-malware software is centrally controlled instead of being
distributed in origin. A better correspondence would be a security
system that would operate on the basis of a peer-to-peer protocol. As
computer viruses and other malware develop, so would defenders that
would be redistributed automatically across the web. Such a
peer-to-peer system would open the door to more opportunities for
malware, but this architecture would give attackers and defenders
equal capabilities, unlike the current situation where attackers have
a wider range of options, with potentially much greater capabilities.

The closest system to the architecture of the immune system from this
perspective is an existing distributed collective response system
where multiple users share signatures of
spam. \cite{DistributedChecksumClearinghouse, Vipul'sRazor} Still,
unless these systems are universally adopted, they do not prevent
widespread infection and thus widespread attack --- i.e. they are not
collective security systems but rather localized security systems.

Second: The current architecture of the Internet is based upon a
protocol (IP \cite{cisco,IPwiki}) that transmits messages independent
of their content. The basic premise is that any valid message is
delivered to its destination. The delivery process involves transfer
from the point of origin through a sequence of nodes (routers) of the
internet. Each router reads the target destination specified by the
message (packet) identifies a node to transfer the message to that
will enable eventual delivery to the destination, and transfers it. In
this proces, there is no evaluation of what the mesage
contains. Individual messages may be lost in transmission due to
network overload, but not due to evaluation of the contents of the
message. This implies that as far as the sender and receiver are
concerned the network is transparent. Any message is transferred from
one node of the network to any other node of the network without
intervention. In considering the transfer of messages, it is important
to recognize that a message is also an action that can be harmful. If
we consider each destination node of the network to be like a
``home,'' and the network to be like the streets, then from the point
of view of security, this is equivalent to having no police on the
streets or military at the borders. Each household, or individual,
must defend him or herself, using means of protection (e.g. guns,
etc.) purchased on the market. That the protection is left to the
individual home reflects the open nature of the Internet. The
corresponding system in a biological sense would enable any cell to
approach and attack any other cell of the body. Under these untenable
conditions there would be no collective security in the medium in
which cells are located. Analogously, there is no protection in the
medium of the Internet.

The two fundamental limitations of the architecture of the Internet
from a security perspective imply that there is no mechanism for a
security system to prevent actions consisting of nodes attacking other
nodes in the Internet. Attacks that are launched cannot be stopped
before they arrive at their destination. There is also no process that
enables removal or elimination of the originators of an
attack. Security systems do not have access to nodes of the Internet
except as they voluntarily participate in security actions by
subscribing to security communities or services. Thus, preventive
action or removal is only possible if the originating node voluntarily
participates in a security action. Without such participation, the
best that can be done is to protect from attack at its
destination. This implies that an intentional attacker can gain
strength and evolve increasingly sophisticated attacks. Attacks can be
focused on particular nodes, either because of their vulnerability or
value. In contrast, to be effective, security must defend all parts of
the system it wishes to protect.

In order to develop an effective collective security system similar to
the immune or human security systems, substantial architectural
changes must be implemented.  Collective security preventing attacks
would require that the routers of the Internet themselves would need
to have protocols that allow refusal of transmission based upon
content or extrinsic information such as point of origin. The routers
of the Internet serve as the transmission medium for the nodes of the
Internet. This corresponds to the intercellular fluids, including the
blood, of physiological systems, the primary locus of immune system
activity. Such an approach was implemented against spam transmission
early in the history of DNSRBLs \cite{ref:DNSRBLs}; however, it
appears to be abandoned. A router based security system would curtail
the ``right of transmission,'' which may be considered fundamental in
discussions of Freedom of speech. \cite{speech}

Absent a router based security system, the second alternative is to
enable automatic transmission of security software among all terminal
(non-router) nodes of the internet. This would enable rapid and
pervasive distribution throughout the system. This is a similar
propagation to that of viruses and other malware. Such automated
transmission might be considered to be less desirable than router
based security, as it involves partial loss of control by owners of
the activities on their computers in favor of security
operations. Corresponding software capabilities exist in peer-to-peer
systems, and in existing voluntary security communities.

Thus far we have not discussed the use of human legal systems to
pursue human originators of malware and spam. In this regard there are
difficulties inherent in international law for pursuing such attacks
as crime. Perhaps more critically, the different domains and time
scales of Internet activity and human legal activity suggest that this
approach will be difficult to utilize to any significant
effect. Criminal prosecution is a high cost and time effort that can
be effective in disrupting non-normative activities but not in
curtailing widespread actions. Indeed, the existing success in
prosecuting Internet crime is limited. \cite{FTCSpam1,FTCSpam2}

\section{Conclusions}

We performed an analysis of principles of security using the immune
system as a model that provides insight into the functioning of any
security system. We identified central functional attributes of a
security system. However, we found that these properties cannot be
effectively implemented in cyber security because of current
limitations of the Internet structure. This is not a limitation of the
security principles but rather of the architecture of the Internet.

Within the current limitations on security systems, security is
largely in the hands of individual nodes. Improving the resistance of
individual nodes to attack is critical because a compromised node will
be the source of attacks on other nodes. The main source of weakness
in the system is the vulnerability of operating systems on individual
computers. Vulnerable operating systems should be avoided not just for
the security of the individual node but for the security of the system
as a whole.

A focus on individual nodes, however, will miss the essential nature
of a collective system and its vulnerabilities that arise due to the
rapid communication.  We therefore recommend that major modifications
will be considered that allow the ability to perform security within
the router systems of the Internet rather than in the end nodes. This
would entail router protocols that can be used to reject messages
based upon content, with distributed detection systems to identify
which content corresponds to harmful acts. Such a security system must
be carefully designed to avoid harm to self through blocking valid
Internet traffic. Indeed, it is the fear of such blocking that has
limited the implementation of such a system.  The use of principles of
security obtained from analysis of other security systems,
specifically the immune system, can provide guidance to enable
effective processes that minimally impede or interfere with valid
traffic.  To this end we summarize our findings of immune system
security.

We found that the three subsystems of the immune system are designed
to address challenges at three scales. The largest scale is the
separation of internal and external domains, i.e. the skin, as well as
other partitions of the system by membranes to separate different
security domains. The intermediate scale corresponds to manifest
system damage, including damage to the boundaries or tissues, that
require repair. The finest scale and highest complexity system
involves responding to distributed collections of molecules or cells
that are difficult to identify in the context of the complex
physiology of the body, and are often able to replicate to create
larger scale damage, e.g. viruses and bacteria.

The most specific and distributed system for security involves
remarkable level of emphasis on detection. Detection can be described
in a general fashion that is relevant to security systems at other
levels of organization including human societal, and cyber
security. Detection involves a standard set of templates that are able
to characterize sufficiently broadly potential invaders, but have
specificity sufficient to distinguish adversaries from self.  Once
detection occurs, multiple processes are used to enhance the ability
to detect the adversaries rapidly and in small numbers.

The detection process is intimately linked to action even though
different agents perform detection and action. The interaction of
action agents and detection agents is a local process involving
identifying individual elements as adversaries and mounting a
proportionate response. These local interactions are carefully
designed to avoid self-inflicted damage.

All of the immune system actions utilize widespread communication and
transportation to transport the detection mechanisms throughout the
system, to recruit additional defenders, and to communicate ongoing
improvements in detection.

A review of the current major systems for cyber security suggests that
there are aspects of cyber security systems that correspond to the
immune system.  However, the fundamental differences in architecture
prevent communication and refinement of responses to adversaries
throughout the system. These constraints also severely limit the
possible counter attacks, including the ability to block sources of
attacks at the points of origin. With such limitations overall success
in suppressing the large numbers of cyber attacks will be
limited. Since the transportation and communication systems can be
exploited by attackers to achieve widespread damage, they must also be
used as part of successful response by defenders.

\vfill\eject

\section{Appendix: Detection and Response in the Immune System}

In this appendix we provide details  about the immune system 
response to identify the mechanisms by which it provides a
capacity for detection and for corrective action.  These are key
aspects of immune system response but are generally not 
currently applicable to cyber security because of the inherent 
architecture of the Internet. As explained in the main paper, 
should the structure of the Internet be modified in a manner
to allow for improved security, these mechanisms can provide
improved guidance for how to proceed.

The adaptive immune system has a high degree of specificity: a
``non-self'' antigen (a signature of an adversary) is recognized and
responses are tailored to that particular pathogen (adversary).  We
will show how the biological system identifies adversary signatures,
transmits information to agents who can take action, determines which
actions are appropriate, and how memory functions.

The cells of the adaptive immune system are known as lymphocytes,
which are special types of white blood cells ({\em leukocytes}).  The
two principal categories of lymphocyte, are {\em B cells} and {\em T
  cells,} are further divided into subcategories as explained below.
Both derive from stem cells found in bone marrow, the {\em
  hematopoietic} or blood-producing cells.\cite{ref:janeway}

B cells have a primary role in identification of adversaries, and T~cells
are the primary agents of action in response to the adversaries. The
tight coupling of identification and action results in mutual regulation
of activity by B and T~cells. When detection occurs, B~cells trigger T~cell
response, and T~cells increase the action of B~cells to accelerate 
detection as necessary. 

\subsection{Detection mechanism}

Detection starts with pattern matching between a detection template and 
entities in the environment that might be characterized as adversaries. 
One type of cell, B~cells, and one type of molecule, antibodies, are the
biological agents primarily responsible for detection of threats. 

The primary template in the adaptive immune system is the antibody.
The antibody is a molecule that binds to anything whose shape matches
(complements) the antibody binding site.  Antibodies can be found in
two primary forms. First, as part of a cell where it is attached on
the surface membrane of a cell to the cell signaling system so that a
match between the antibody and something in the environment triggers a
cell response. Second, antibodies can be released into the blood and
when floating around attach themselves to matching entities.

The key to the use of the antibody as a pattern matching template is
the possibility of generating many different shapes of the molecule
binding site. This ability arises because the antibody is formed of
several parts, the parts can be combined together in flexible ways,
and the parts themselves can be varied.  An antibody is comprised of
four peptide chains arranged to form a Y-shape. Each of the Y-shape's
two ``arms'' has a variable region.\cite{ref:alberts}

The matching between an antibody through binding to another molecule
(called an antigen) occurs at the molecular level: the shape of an
antibody is complementary to that of the antigen to which it
binds. More precisely, it will bind to a part of the antigen called
the {\em epitope,} typically about 600 \AA$^2$ in
area.\cite{ref:perelson1997}

Antibodies are generated internally by B~cells. The mechanisms for
their manufacture are through the read-out of genetic templates for
each of the components and their combination into a single
molecule. B~cells also are designed to carry the antibodies they
generate on surface membrane receptors as a detection mode.  When the
antibody binds to something in the environment the B~cell can change
its behavior accordingly.

A B~cell identifies a threat when the antibodies present on its
surface bind to a specific foreign antigen.\cite{ref:sproul}
B~cells that have identified an adversary by binding to it 
perform a series of actions to improve future detection of
similar adversaries and initiate response. The actions they 
take to initiate response will be discussed in the next section.

\subsection{Detection templates}

B cells determine which templates to use for detection of invaders
through a process which requires multiple steps, each of which is
central to the functioning of the immune system.

The first step is that multiple templates can be generated by cellular
mechanisms.  By recombining genetic elements from a ``toolbox'' of
short sequences, a large volume in the shape space of all possible
antigens can be covered.

The second is the process of generating the set of templates that are
found in the body as B~cells are formed, mostly in bone marrow.
B~cells are produced from a type of stem cell. They progress through
several developmental stages marked by changes in the genetic
sequences which define the active sites of their antibody molecules.
This variation during their development gives rise to a remarkable
diversity of templates associated with distinct B~cells.  This large
array of prototype templates don't exhaustively cover the set of
possibilities, but are widely dispersed throughout the set of
possibilities.

The third is eliminating templates that correspond to self molecules.
Mounting an immune response to the body's own molecules and cells is
extremely undesirable. The body avoids this (not always perfectly) by
inactivating lymphocytes which happen to be self-reactive after they
are generated. During the last stages of their maturation, B cells are
presented with self molecules.  B cells which react to these molecules
are triggered into programmed cell death (apoptosis) or modifying
their receptors. \cite{ref:alberts} This process occurs in part after
developing B cells move to the spleen, where they complete their
maturation process.

The fourth is enhancing the presence of B~cells whose antibodies have
responded in the past to antigens. This serves as a form of
memory. Cells that have not been exposed to antigen develop into
different forms once they have encountered their target antigen.  Some
become {\em effector cells,} which go on to play an active role in
immune response; others become {\em memory cells,} which do not
immediately participate in anti-pathogen defense but are sensitized to
an antigen so that a second encounter will stimulate them to become
effector cells.  Over the short term their stimulation by antigens
that are present in the current attack contributes to the strength of
the response. Memory cells are much longer-lived than effector cells,
persisting sometimes until the death of the animal instead of dying
within days, so they can facilitate rapid immune responses to threats
which have been encountered before.

The fifth is the enhancement of templates matching to current invaders
by the process of evolution, discussed in the next section.

The sixth is the communication between B~cells that enables multiple
distinct antibodies to be formed to the same adversary. This is
discussed below in the section on B-cell to B-cell communications.

The seventh is the use of indicators of cellular damage that may
reduce the threshold for detection or trigger it. \cite{ref:matzinger}

\subsection{Evolution of detection}

When a body is attacked by a reproducing agent, such as a virus 
or a bacterium, the rate of detection of that invader is critical to
the success of the response. Detection of even individual ones
is necessary before they can reproduce, requiring remarkably
fast and pervasive detection. The detection 
can be enhanced if the detection template is improved to be a
better match to the invader. This is accomplished by a process
of progressive improvement through selecting improved versions,
i.e. evolution.

The evolutionary process occurs primarily in specific organs of
the body called lymph nodes. \cite{ref:pierre} When B~cells detect an invader
with their antibodies, they take the piece of the invader that
binds to the antibody. They travel to a lymph node and place the
antigen in the wall of the lymph node. B~cells reproduce very
rapidly in a lymph node. In doing so they vary by high
rates of mutation their own DNA which codes for the antibody
they produce. They use their antibody to test its binding to the
antigens in the wall of the lymph node.  The less successful
ones rapidly die by programmed cell death.  The ones that are
most successful, i.e. the best detectors, are sent out into the 
blood stream to engage with the adversary. 

As B~cells bring pieces of the
invaders into the lymph nodes, and the process of selecting
improved binders continues, the B~cells released back into
the blood stream are continually improved in their ability to
detect the invader. This leads to a dramatic change in the
sensitivity of the immune system to that invader, enabling the
system to completely eliminate the invader.

\subsection{B-cell to B-cell communication: Improving detection}

Once an adversary has been detected, identifying multiple
ways to detect it both increases the ability of detection
and reduces the possibility of an adversary acting to avoid detection
by modifying their structure and behavior.

B cells that have identified an invader by binding to it, use this
detection to enhance the ability of other cells to identify the invader.
The method for doing this is to make less specific parts of the invader
to which other B~cells may be able to bind. The B~cell
breaks parts of the invader into smaller components and displays
these components on the B~cell surface. Other B~cells then bind to
these molecules triggering their own response systems. This creates 
additional methods for invader detection in addition to the original 
one detected by the first B~cell. 

Since adversaries are not individual molecules or cells, but rather
are replicating viruses and bacteria, the immune system utilizes 
massive parallelism:  many types of pattern detection templates 
(antibodies) can be tested simultaneously across the system. This
enables using multiple methods to detect the same type of adversary.

The communication by B~cells to other B~cells in this way is a key form 
of  cooperation between cells of the adaptive immune system. The
network of interactions also includes communication 
from B~cells to T~cells, and reciprocally from T~cells to B~cells.

\subsection{B-cell to T-cell communication and intelligence}

Cell to cell communication enables B~cells to communicate the discovery
of antibodies to the cells responsible for action --- T~cells.  Cell to cell 
communication is also important for information gathering by T~cells 
from non-immune cells --- the analog of human intelligence. 

T~cells use a particular ``secure" communication channel to receive messages
from B~cells and other cells of the body. The security of the communication is
provided by MHC molecules. These are a diverse set of molecules that are
highly specific to each individual. The specific nature of these molecules is
maintained by high genetic variability of the MHC molecules over evolutionary
time and within the human population at any one time. Secure communications 
are important to prevent invaders from ``spoofing" messages to the T~cells.

When a B~cell communicates to T~cells, the ``information," is in the form of a 
molecular fragment on the surface of the B~cell bound to a MHC molecule. 
T~cells have the ability to bind to and recognize the MHC molecule and the 
molecular fragment that is bound to it.

As described before, a B cells whose antibodies bind to something in the
environment (the antigen) takes the antigen/antibody complex and breaks 
it into small units, which are returned to the surface. At the surface the
fragments are combined with a MHC molecule. 
The surface display of the MHC with the parts of the antigen/antibody
complex are then detected by T~cells in the vicinity triggering T~cell
action. 

Similarly, any cell of the body may display MHC molecules with fragments
of molecules that are found inside the cell. This enables T~cells to check
for normalcy or cell infection (e.g. by viruses) from the surface display on
a cell.

\subsection{Action agents: T~cells and phagocytes}

The primary agents responsible for action against threats in the immune 
system are T~cells. T~cells both secrete molecular messengers which can 
kill other cells or can increase the growth rate of other cells.

T~cells can secrete cytotoxins which can kill other cells. 
T~cells target foreign cells or self cells that have been infected by 
foreign agents such as viruses. T~cell action is triggered by
their binding to complexes of antibody and MHC molecule, i.e. they
are triggered to action by B~cell signaling.

T~cells also are responsible for signaling B~cells into a more active
state.  This active state results in two effects. The first is rapid
growth and proliferation.  The second is the manufacturing of more
antibodies and their release into the blood and other intercellular
fluids so that they bind to antigens throughout the system and not
just when a B~cell encounters them. This is a process that makes the
detection itself a form of action. The binding of the antibodies can
disable the pathogens directly or mark them for destruction.  The
former is accomplished by interfering with the surface molecules which
pathogens use to infect cells or by binding to the toxins produced by
bacteria.

The two different functions of T cells are divided between killer T cells
and helper T cells. The dynamics of killer and helper T-cell response
is regulated in part by their responsiveness to distinct MHC molecules.

Finally, a method is necessary to clean up both the dead cells and 
molecules, and even live but marked cells, from the system. This is 
accomplished by phagocytes. Phagocytes, are cells that can engulf 
and use digestive juices to, in effect, eat other cells and molecules. 
Their action is triggered by the presence of the antibodies bound
to molecules and cells after their release by B~cells. \cite{ref:alberts}

\subsection{Post response: Reducing the cohort and memory}

The rapid growth of the number of B~cells, antigens and T~cells in
response to an adversary also requires a mechanism to reduce that
number when the response is successful. As the number of antigens
declines, the short lifespan of the B~cells and the absence of
triggers to produce additional ones leads to their decline. Additional
triggers to accelerate this process may also take place but must be
carefully designed to avoid exploitation.

The immune system is costly, in that providing an immunological 
defense against foreign pathogens requires resources --- such as energy 
and nutrients --- which are also needed for muscular and neural activities, 
reproduction, growth and other biological processes.\cite{ref:lochmiller}

Still, once a specific threat has been identified, it is more likely
to reappear. This justifies retention of identification templates,
in antigen and B~cell form, for future use. This retention accelerates 
the response when the same threat reappears. 

Some B~cells preserve the memory of prior responses by continuing
to survive for extended periods of time after a response.  Since some
death is unavoidable, they also replicate without mutation to preserve 
their antibodies in their offspring. These ``Memory cells'' maintain the 
capacity to generate their specialized responses, so that a later infection 
by the same pathogen can be countered quickly.  

Organisms without an adaptive immune system lack such immunological
memory, whereby a pathogen can be ``remembered'' by its signature
antigens.

\subsection{Security failures}

The immune system, like other systems, has modes of failure that
demonstrate how its processes work and fail to work in addressing key
challenges.

The first type of failure is when an invading organism cannot be
overcome by the immune system. Bacteria are significant challenges to
the immune system, and before antibiotics were available many people
died due to failure of the immune system to respond sufficiently
effectively to bacterial diseases. Such failures reflect the tight
competitive balance between bacterial attackers and immune system
response. When the immune system is weakened due to lack of nutrition,
compound diseases or particularly effective pathogenic attackers the
immune system response may not be sufficient to prevent growth and
replication of the bacteria.

The second type of failure is auto-immune disease. This is a failure
to distinguish self from non-self which causes the immune system to
attack self cells. This is most common for the case of some special
types of cells, such as insulin-producing cells. It is believed that
the small number of such cells leads to their vulnerability as the
immune system may not sufficiently be exposed to these cells and
therefore is more likely to identify them as ``other'' and attack them,
thereby giving rise to some types of diabetes.\cite{ref:pozzilli}

The significance of both of these failure modes for other forms of
security is that even with the best system possible, success is not
guaranteed.

\vfill\eject

\end{document}